\documentclass[a4paper,fleqn]{cas-dc}

\usepackage[numbers,compress,square,comma]{natbib}	

\def\tsc#1{\csdef{#1}{\textsc{\lowercase{#1}}\xspace}}
\tsc{WGM}
\tsc{QE}

\usepackage{tikz}
\usetikzlibrary{calc}
 \usetikzlibrary{decorations.text}
 \usetikzlibrary{shapes}
 \usetikzlibrary{decorations.pathmorphing}
\usetikzlibrary{decorations.pathreplacing}
\usetikzlibrary{arrows.meta}
\tikzset{
  >={To[length=5pt]}
  }
\usetikzlibrary{shapes, shapes.geometric, shapes.symbols, shapes.arrows, shapes.multipart, shapes.callouts, shapes.misc}
\newcommand{\mz}{\mathcal{Z}}
\newcommand{\mf}{\mathcal{F}}

\DeclareMathOperator{\R}{\mathbb{R}}

\newcommand{\dd}{\mathrm{d}}

\begin{document}
\let\WriteBookmarks\relax
\def\floatpagepagefraction{1}
\def\textpagefraction{.001}

\shorttitle{Massive flows in \texorpdfstring{AdS$_6$/CFT$_5$}{AdS6/CFT5}}    

\shortauthors{}  

\title [mode = title]{Massive flows in \texorpdfstring{AdS$_6$/CFT$_5$}{AdS6/CFT5}}  



%

\author[1]{Mohammad Akhond}


\ead{akhond@yukawa.kyoto-u.ac.jp}

\affiliation[1]{organization={Yukawa Institute for Theoretical Physics, Kyoto University},
            city={Kyoto},
            postcode={606-8502}, 
            country={Japan}}

\author[2,3]{Andrea Legramandi}

\ead{andrea.legramandi@unitn.it}

\affiliation[2]{organization={Pitaevskii BEC Center, CNR-INO and Dipartimento di Fisica, Universit\'a di Trento},
            city={Trento},
            postcode={38123}, 
            country={Italy}}
            
\affiliation[3]{organization={Department of Physics, Swansea University},
            city={Swansea},
            postcode={SA2 8PP}, 
            country={United Kingdom}}

\author[3]{Carlos Nunez}

\ead{c.nunez@swansea.ac.uk}

\author[4]{Leonardo Santilli}

\ead{santilli@tsinghua.edu.cn}

\affiliation[4]{organization={Yau Mathematical Sciences Center, Tsinghua University},
            city={Beijing},
            postcode={100084}, 
            country={China}}

\author[3]{Lucas Schepers}

\ead{988532@Swansea.ac.uk}

\begin{abstract}
We study five-dimensional ${\cal N}=1$ Superconformal Field Theories of the linear quiver type. These are deformed by a relevant operator, corresponding to a homogeneous mass term for certain matter fields. The free energy is calculated at arbitrary values of the mass parameter. After a careful regularisation procedure, the result can be put in correspondence with a calculation in the supergravity dual background. The F-theorem is verified for these flows, both in field theory and in supergravity. This letter presents some of the results in the companion paper \cite{Akhond:2022oaf}.
\end{abstract}

\maketitle

\section{Introduction}

After Nahm's famous classification of superalgebras containing the conformal symmetry \cite{Nahm:1977tg}, the pursuit  and study of superconformal field theories (SCFTs) in higher dimensions has completely reshaped the quantum field theory (QFT) landscape. Five-dimensional $\mathcal{N}=1$ SCFTs \cite{Seiberg:1996bd} are the main characters of this work. They  do not admit a weakly coupled description \cite{Cordova:2016xhm,Chang:2018xmx}, thus remaining out of the reach of the textbook Lagrangian formalism. Developing new methods to study them, based on string \cite{Aharony:1997bh} or M-theory \cite{Intriligator:1997pq}, turned out to be an invaluable source of lessons, exportable to QFT in general.\par
Supersymmetric localisation \cite{Pestun:2007rz} provides a handle on the problem of computing physical observables in strongly coupled SCFTs. The basic idea is to work with quantities that are protected by supersymmetry, such as the sphere partition function. These quantities are amenable to be analysed away from the strongly coupled superconformal point. Deforming the SCFT into one of its gauge theory phases and applying the localisation machinery, the path integral reduces to an ordinary integral. Taking the strong coupling limit only at the end of the computation sheds light on the properties of the SCFT itself.\par
Combining the vast progress in five-dimensional SCFTs with the AdS/CFT correspondence \cite{Maldacena:1997re} opens up a plethora of avenues to explore. The computational strategy just outlined has been fruitfully employed in \cite{Jafferis:2012iv,Alday:2014rxa,Alday:2014bta,Chang:2017mxc} to perform refined checks of the duality between certain 5d $\mathcal{N}=1$ SCFTs and warped AdS$_6 \times \mathbb{S}^4$ geometries \cite{Romans:1985tw,Cvetic:1999un} and orbifolds thereof \cite{Bergman:2012kr} in massive Type IIA supergravity. More recently, warped AdS$_6 \times \mathbb{S}^2 \times \Sigma $ backgrounds have been constructed in Type IIB string theory \cite{DHoker:2016ujz,DHoker:2016ysh,DHoker:2017mds,DHoker:2017zwj}, and proposed to be dual to SCFTs described by balanced linear quivers. Also in this case, field-theoretical computations based on localisation provided support for the duality \cite{Gutperle:2018axv,Gutperle:2018vdd,Fluder:2018chf,Uhlemann:2019ypp,Uhlemann:2020bek}.\par
The aim of this letter is to delve further into the subject. We consider ${\cal N}=1$ linear quivers and their dual supergravity backgrounds. We analyse the effect of turning on relevant deformations in the SCFT. These are mass deformations that trigger a Renormalization Group (RG) flow. We  study the flows between SCFTs in the same class both at the beginning and at the end of the flow. Our goal is to probe the AdS$_6$/CFT$_5$ correspondence both at and away from the SCFT points. The main achievement is to find holographic agreement at every point of the RG flow, for the class of models and  relevant deformations that we consider. In particular, we present strong evidence for the existence of an F-theorem in these flows.\par
The details of the RG flows are spelled out in Section \ref{sec:qft}, and schematically presented in Fig. \ref{fig:flow}. Then, we compute the sphere free energy for arbitrary values of the mass parameter and elaborate on a 5d F-theorem. Along the way, we encounter an anomaly first discussed in \cite{Chang:2017cdx}. A careful treatment of it is essential to ensure the validity of the AdS/CFT dictionary.\par
In Section \ref{sec:gravity} we present the solutions for the supergravity background and their relation to an electrostatic problem. This relation was thoroughly examined in \cite{Legramandi:2021uds}, and here we extend the analogy by including a natural deformation of the setup. We then identify this deformation as the holographic dual to the relevant operator added to the dual SCFT, and find that the supergravity background and the large $N$ sphere free energy are described by the same electrostatic picture. Furthermore, we compute the holographic central charge (i.e. the quantity dual to the 5d free energy), again showing perfect agreement with the field theory result. For more details on the derivations, as well as extensions in several directions, including the analogous AdS$_4$/CFT$_3$ setting, we refer to the companion work \cite{Akhond:2022oaf}.

\section{Mass deformations of 5d long quivers}
\label{sec:qft}
We consider 5d SCFTs that admit a gauge theory phase with gauge group $\prod_{j=1}^{P-1} SU(N_j)$ and matter fields in the bifundamental of $SU(N_j) \times SU(N_{j+1})$, plus additional $F_j$ flavours in the fundamental of $SU(N_j)$. The data of each such theory are encoded in a linear quiver:
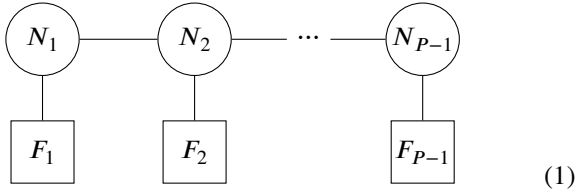
\begin{equation}
\label{fig:quiverfiga}
	\begin{tikzpicture}[auto,square/.style={regular polygon,regular polygon sides=4}]
						\node[circle,draw] (gauge1) at (2,0) { \hspace{20pt} };
						\node (a1) at (2,0) {$N_{P-1}$};
						\node[draw=none] (gaugemid) at (0.5,0) {$\cdots$};
						\node[circle,draw] (gauge3) at (-1,0) { \hspace{20pt} };
						\node[circle,draw] (gauge4) at (-3,0) { \hspace{20pt} };
						\node (a2) at (-3,0) {$N_{1}$};
						\node (a3) at (-1,0) {$N_{2}$};
						\node[square,draw] (fl1) at (2,-1.5) { \hspace{10pt} };
						\node[square,draw] (fl2) at (-3,-1.5) { \hspace{10pt} };
						\node[square,draw] (fl3) at (-1,-1.5) { \hspace{10pt} };
						\node[draw=none] (aux1) at (2,-1.5) {$F_{P-1}$};
						\node[draw=none] (aux2) at (-3,-1.5) {$F_1$};
						\node[draw=none] (aux3) at (-1,-1.5) {$F_2$};
						\draw[-](gauge1)--(gaugemid);
						\draw[-](gaugemid)--(gauge3);
						\draw[-](gauge4)--(gauge3);
						\draw[-](gauge1)--(fl1);
						\draw[-](gauge4)--(fl2);
						\draw[-](gauge3)--(fl3);
					\end{tikzpicture}
\end{equation}
We restrict our attention to balanced quivers, subject to 
\begin{equation}\label{balance}
	F_j = 2N_j - N_{j-1} - N_{j+1} .
\end{equation}
Supergravity backgrounds for this class of theories have been constructed in \cite{DHoker:2016ujz,DHoker:2016ysh,DHoker:2017mds,DHoker:2017zwj,Legramandi:2021uds}.\par
Balanced linear quivers are characterised by a rank function $\mathcal{R} (\eta)$. It is a piecewise-linear function of the continuous variable $0 \le \eta \le P$ with $\mathcal{R}(j)=N_j$. The number of flavours $F_j$ is related to the discontinuity of $\partial_{\eta} \mathcal{R} $ at $\eta=j$ via \eqref{balance}.\par
The theories \eqref{fig:quiverfiga} have rich moduli spaces of vacua. The Coulomb branch (CB) is fibered over the parameter space of real masses for the matter fields. When all flavours are massless, the CB is a cone, and the SCFT is located at its tip. However, giving mass to a subset of the flavours, the conical singularity is smoothed into a less severe one.\par
In this letter we explore the effect of turning on mass deformations for balanced linear quivers. We select a homogeneous deformation, in which many flavours at each node acquire equal mass $m$, so that there is only one mass scale in the problem.\footnote{The generalisation to an arbitrary number of masses is reported in \cite{Akhond:2022oaf}.} The choice of relevant deformation of the SCFT is specified by selecting the number $F_{2,j}$ of flavours that are given a mass, at each $j$, while the remaining $F_{1,j}=F_j - F_{2,j}$ stay massless. Condition \eqref{balance} then selects a splitting $N_j =N_{1,j}+N_{2,j}$. In other words, a relevant deformation of the SCFT which is controlled by a unique parameter $m$ and such that \eqref{balance} holds at the end of the RG flow induced by the deformation, is specified by a splitting of the rank function $\mathcal{R} (\eta)$ into
\begin{equation}
\label{R1R2eqR}
	\mathcal{R}_1 (\eta) + \mathcal{R}_2 (\eta) = \mathcal{R} (\eta) .
\end{equation}
For later reference, we write the Fourier expansion,
\begin{equation}\label{eq:doublefourier}
    \mathcal{R}_{\alpha} (\eta) = \sum_{k=1}^\infty R_{\alpha,k}\sin\left(\frac{k\pi \eta}{P}\right) , \qquad \alpha=1,2
\end{equation}
with $R_{1,k} + R_{2,k} $ subject to \eqref{R1R2eqR}.\par
\begin{figure}[t]
\centering
\begin{tikzpicture}[scale=0.75]
\node[] (t1) at (0,4.5) {$\bullet$};
\node[anchor=east] (cft1) at (t1.west) {{\small SCFT$[\mathcal{R}]$}};
\node[] (t2) at (3.5,2) {$\bullet$};
\node[align=left,anchor=north west] (cft2) at (t2.south) {{\small SCFT$[\mathcal{R}_1] \ \oplus $ SCFT$[\mathcal{R}_2]$}\\ {\small $\oplus$ Massive fields}};
\path[->,black] (t1) edge node[anchor=south west,pos=0.8] {${\scriptstyle m \to \infty}$} (t2);
\node[anchor=west] (m0) at (0.5,4.3) {${\scriptstyle m \to 0 }$};

\node[anchor=east] (uv) at (-2,5.5) {\small UV};
\node[anchor=east] (ir) at (-2,0.5) {\small IR};
\draw[->,thick] (ir) -- (uv);

\end{tikzpicture}
\caption{RG flow activated by the relevant deformation which gives equal mass to a large number of flavours across the quiver. The UV SCFT with rank function $\mathcal{R}$ flows to two SCFTs with rank functions $\mathcal{R}_1$ and $\mathcal{R}_2$, plus a collection of heavy fields that decouple at the very end of the flow.}
\label{fig:flow}
\end{figure}
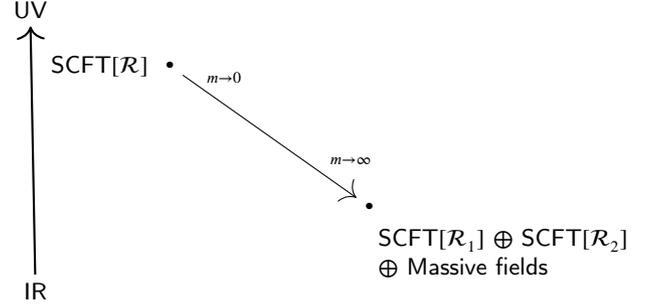
Turning on $m>0$ activates an RG flow away from the SCFT characterised by $\mathcal{R}$. We show that at the end of the flow we find two factorised balanced SCFTs, with rank functions $\mathcal{R}_1$ and $\mathcal{R}_2$. In addition, there is a collection of fields that have acquired a large mass along the flow and have eventually decoupled. The situation is represented in Fig.~\ref{fig:flow}. Concretely, these massive fields account for Higgsed W-bosons and matter modes that are present in the original SCFT but are missing in the pair of SCFTs at the end of the flow. Indeed, the breaking $SU(N_j) \to SU(N_{1,j}) \times SU(N_{2,j})$ gives mass to some of the W-bosons. Besides, out of the $F_j N_j$ flavour zero-modes, only $F_{1,j} N_{1,j} + F_{2,j} N_{2,j}$ remain, the other ones having decoupled, and likewise for the bifundamental fields.\par
As we have mentioned, the original UV SCFT sits at the gapless vacuum, on the singularity at the origin of the CB. A finite mass splits the singularity in two, separated by a finite distance $m$. In the limit $m \to \infty$, the CB will eventually break in two disconnected components, each with its own single gapless vacuum yielding a 5d SCFT. From the IR perspective, each such vacuum is blind to the other CB, and the full RG flow is uniquely specified by the choice of which CB conical singularity is retained and which is sent to infinity. The sphere free energy, however, is computed in the UV and, as we  will show, it retains the information about the factorisation of the CB. In particular, it is independent of the choice of vacuum, the various choices being related by a change of variables in the matrix model \eqref{Z}. With this caveat in mind, we loosely refer to the situation in Fig.~\ref{fig:flow} as the RG flow.

Massive deformations of this type have a neat interpretation in Type IIB string theory. Given a five-brane web engineering the SCFT \eqref{fig:quiverfiga}, we separate the D5 branes in two stacks and pull them apart a distance $m$. Modes arising from strings stretched between branes in the same stack remain massless, while strings stretched between the two stacks are heavy.\par
In order to quantitatively probe the AdS/CFT correspondence under such deformations, we now move on to discuss the large $N$ limit of the free energy on $\mathbb{S}^5$.

\subsection{Long quiver free energies}
The partition function of the linear quiver on $\mathbb{S}^5$ is computed by a matrix model. It takes the form \cite{Kallen:2012va,Lockhart:2012vp}
\begin{equation}
\label{Z}
	\mz_{\mathbb{S}^5} = \frac{1}{N!} \int \dd \vec{\phi} ~ e^{-S_{\text{eff}} (\vec{\phi}, m) }   Z_{\text{non-pert}} .
\end{equation}
where $\vec{\phi}$ collectively denotes the CB parameters, the integration domain is $\mathbb{R}^{\sum_j N_j} $ and the measure $\dd \vec{\phi}$ includes the traceless constraint at each gauge node. The matrix model effective action $S_{\text{eff}}$ contains both the classical action as well as the one-loop contributions from W-bosons and fundamental and bifundamental matter fields. We refer to \cite{Kallen:2012va} for the explicit expressions. The dependence on the mass $m$ appears through a shift $\phi_j \mapsto \phi_j + m$ in the argument of the one-loop determinant of the $F_{2,j}$ flavours at each $j$. Finally, $ Z_{\text{non-pert}} \to 1$ in the large $N$ limit, hence we neglect it from now. In addition, one may enrich \eqref{Z} with contributions from the background fields, considered in \cite{Chang:2017cdx,Chang:2019uag}.\par
The holographic regime corresponds to choose an integer $N$ and take $N \gg 1$ with $N_j/N$ fixed. Moreover, we take a large $N$ limit in which also the length $P$ of the quiver grows. This long quiver limit was first addressed in \cite{Uhlemann:2019ypp} and we extend it to the mass deformed setup. This will allow us to evaluate the free energy at every point of the flow.\par
As usual, the effective action is large when $N \gg 1$, thus the leading contribution to \eqref{Z} comes from the saddle points of $S_{\text{eff}}$. The interplay with $P \gg 1$ introduces novel features compared to the large $N$ limit familiar from, for example, $\mathcal{N}=4$ super-Yang--Mills in 4d. Expanding on \cite{Uhlemann:2019ypp}, we are led to make the scaling ansatz 
\begin{equation}
\label{scalingansatz}
	\phi , m \propto P^{\chi} 
\end{equation}
for some $\chi >0$. That is, in the holographic regime ($P$ very large), the CB parameters as well as the mass grow with $P$. With this assumption, the functions appearing in $S_{\text{eff}}$ have large argument, and the physically realised value of $\chi$ is determined by the existence of an equilibrium configuration. Self-consistency of the scaling limit tells us that $\chi=1$.\par
The procedure to study the large $N$ limit is detailed in \cite{Akhond:2022oaf}, building on \cite{Uhlemann:2019ypp}. With the long quiver scaling \eqref{scalingansatz}, we introduce the eigenvalue density $\rho (\eta, \phi)$ \cite{Uhlemann:2019ypp}. At every fixed $\eta=j$ we have the usual eigenvalue densities, and this function $\rho(\eta, \phi)$ collects them together. In the regime $N,P \gg 1$ we have \cite{Akhond:2022oaf} 
\begin{align}
	S_{\text{eff}} &= - \frac{\pi}{6} N^2 \int_0 ^P \dd \eta \int \dd \phi \rho (\eta,\phi) \label{SeffLargeN} \\
	& \left[  - \frac{1}{N} \left( \partial_{\eta} ^2 \mathcal{R}_1 (\eta) \lvert \phi \rvert^3  + \partial_{\eta} ^2 \mathcal{R}_2 (\eta)  \lvert \phi -m\rvert^3 \right) \right. \notag \\
		&  \left. + \int \dd \tilde{\phi} \left( \frac{27}{4} \rho (\eta,\tilde{\phi}) \lvert \phi - \tilde{\phi} \rvert+ \frac{1}{2} \partial_\eta ^2 \rho (\eta,\tilde{\phi}) \lvert  \phi - \tilde{\phi} \rvert^3 \right) \right] . \notag 
\end{align}
Notice that $\mathcal{R}_{1,2}$ is linear in $N$, thus the $1/N$ in the first line guarantees that all terms are of the same order and contribute to the saddle point. Extremising \eqref{SeffLargeN} and acting with $\partial_{\phi}^3$, we arrive at the saddle point equation \cite{Akhond:2022oaf}
\begin{equation}\label{spe}
\begin{aligned}
	& \frac{9}{4} \partial_{\phi} ^2 \rho (\eta, \phi) + \partial_{\eta} ^2 \rho (\eta, \phi) \\
	&- \frac{1}{N}  \partial_{\eta}^2 \left[ \mathcal{R}_1 (\eta)  \delta \left( \phi\right)+ \mathcal{R}_2 (\eta)  \delta \left( \phi-m\right) \right] =0 .
\end{aligned}
\end{equation}
This is a Poisson equation for $\rho$, with sources at $\phi=0$ and $\phi=m$. It describes an electrostatic problem, which we will encounter again in the study of the supergravity dual. The dictionary  between QFT and supergravity is spelled out in \eqref{holotocft} below.\par
Solving \eqref{spe}, we use $\rho(\eta, \phi)$ to compute the free energy $\mf = \ln \mz_{\mathbb{S}^5} $. The result is \cite{Akhond:2022oaf}:
\begin{equation}
\label{F}
\begin{aligned}
	\mf = \frac{27}{32}P^2 &\sum_{k=1}^{\infty}\frac{1}{k} \left[ R_{1,k}^2 + R_{2,k}^2 +2 R_{1,k} R_{2,k}e^{-\frac{2\pi k m}{3P} } \right] \\
	+ &\sum_{k=1}^{\infty}\frac{1}{k} R_{1,k} R_{2,k} \left[ \frac{9\pi}{8}P k m +  \frac{\pi^3}{12P} k^3 m^3 \right] .
\end{aligned}
\end{equation}
We recognise in the first two pieces the free energies of two SCFTs with rank functions $\mathcal{R}_1$ and $\mathcal{R}_2$, respectively. The $m$-dependent terms describe an interaction between the two quivers. All the terms are $O(P^2)$, according to \eqref{scalingansatz}.\par
It is instructive to analyse the cubic part in $m$ in the second line of \eqref{F}, which we denote $\delta \mf$. Using the relation between $\partial_{\eta}^2 \mathcal{R}_2 \vert_{\eta=j}$ and $F_{2,j}$, it can be conveniently recast in the form \cite{Akhond:2022oaf}
\begin{equation}
	\delta \mf = \sum_{j=1}^{P-1} \frac{\pi}{3} k^{\text{eff}}_{f,j} ~ \mathrm{Tr} \left(M_j ^3 \right) , \qquad k^{\text{eff}}_{f,j} = \frac{\mathcal{R}_1 (j)}{2} ,
\end{equation}
where $M_j$ is the mass matrix whose eigenvalues are $0$ with multiplicity $F_{1,j}$ and $m$ with multiplicity $F_{2,j}$. Now recall that, upon localization, a supersymmetric Chern--Simons (CS) term in 5d is 
\begin{equation}
\label{CSbg}
	S_{\text{CS}} = \frac{\pi k}{3} \mathrm{Tr} \left( M^3 \right) 
\end{equation}
where $M$ is the scalar superpartner of the gauge field. We take it to be a background field for the flavour symmetry, rather than a dynamical field. As usual, gauge invariance of \eqref{CSbg} requires $k \in \mathbb{Z}$. As it only involves background fields, we are free to add the counterterm \eqref{CSbg} to our action. We observe that $\delta \mf$ can be cancelled by a sum of such local, supersymmetric counterterms, with \emph{fractional} CS couplings $- k^{\text{eff}}_{f,j} \in \frac{1}{2} \mathbb{Z}$. The non-integrality of the CS couplings spoils invariance under background large gauge transformations and signals the presence of an anomaly (see \cite{Chang:2017cdx} for in-depth discussion). We include such counterterm and denote $\tilde{\mf}$ the improved free energy. Let us mention that a similar dilemma is faced in matching the rank-$N$ $E_n$ theories with their holographic dual in Romans $F(4)$ supergravity \cite{Chang:2017mxc}.\par

\subsection{F-theorem}
We now introduce the quantity 
\begin{equation}
\label{FEFTdef}
	\mf_{\text{EFT}} = \tilde{\mf} - \mf_{\text{dec.}} 
\end{equation}
where $\tilde{\mf}$ is the free energy improved with the background CS counterterm. From this free energy of the mass-deformed quiver with rank function $\mathcal{R}$, we subtract the contribution $\mf_{\text{dec.}}$ from the fields that become heavy and eventually decouple. We claim that $\mf_{\text{EFT}}$ is the proper quantity to explore the RG flow between the original SCFT and the two decoupled SCFTs. The underlying motivation to consider \eqref{FEFTdef} is that usually one wants to make statements that are intrinsic to the interacting SCFT, regardless of free massive fields that may have originated. In fact, both holography \cite{Maldacena:1997re} and the F-theorem \cite{Klebanov:2011gs} are claims of this sort.\par
Concretely, the following statements hold:
\begin{itemize}
\item $\mf_{\text{EFT}}$ satisfies the 5d F-theorem;
\item $\mf_{\text{EFT}}$ matches with the holographic central charge computed in supergravity.
\end{itemize}
We leave the technical evaluation of $\mf_{\text{EFT}}$ to the companion work \cite{Akhond:2022oaf}. It is based on analysing which flavours and W-bosons become massive along the flow and computing their free energy, denoted $\mf_{\text{dec.}}$, based on an Effective Field Theory (EFT) approach. More explicitly, we count the total number of such heavy modes that become decoupled at the end of the RG flow, and evaluate their free energy. Crucially, the EFT approach consists in cutting off energy scales larger than $m$.\footnote{The background Chern--Simons term is regularised in the same vein.} Doing this exercise for the heavy W-bosons and matter fields and adding all together, we obtain $\mf_{\text{dec.}}$. We then subtract this quantity from \eqref{F}.
The result is,
\begin{equation}
\label{FEFTsolved}
\begin{aligned}
	\mf_{\text{EFT}} [\mathcal{R}_1, \mathcal{R}_2]  &= \frac{27}{32}P^2 \sum_{k=1}^{\infty}\frac{1}{k} \left[ R_{1,k}^2 + R_{2,k}^2 \right. \\
		& \left. +2 R_{1,k} R_{2,k}e^{-\frac{2\pi k m}{3P} } \left( 1+ \frac{2\pi k m}{3P} \right) \right] .
\end{aligned}
\end{equation}
Denoting $\mf \left[ \mathcal{R}\right]$ the free energy of a conformal (massless) quiver with rank function $\mathcal{R}$, we have that \eqref{FEFTsolved} satisfies
\begin{equation}
\begin{aligned}
	&\lim_{m/P \to 0} \mf_{\text{EFT}}\left[ \mathcal{R}_1, \mathcal{R}_2 \right]  = \mf \left[ \mathcal{R}_1 + \mathcal{R}_2  \right] , \\
	&\lim_{m/P \to \infty} \mf_{\text{EFT}} \left[ \mathcal{R}_1, \mathcal{R}_2 \right] = \mf \left[ \mathcal{R}_1 \right] + \mf \left[ \mathcal{R}_2 \right]  \\
	&\partial_{m} \mf_{\text{EFT}}  \le 0 , \quad  \lim_{m/P \to 0} \partial_{m} \mf_{\text{EFT}} = 0 = \lim_{m/P \to \infty} \partial_{m} \mf_{\text{EFT}} .
\end{aligned}
\label{Fthm}
\end{equation}
In words: $\mf_{\text{EFT}} $ is monotonically decreasing, stationary at the SCFT points, and interpolates between the free energies of the UV SCFT, with rank function $\mathcal{R}=\mathcal{R}_1 + \mathcal{R}_2 $, and of the IR SCFT, comprised of the two quivers with ranks functions $\mathcal{R}_1$ and $\mathcal{R}_2$. These are precisely the requirements of the strong version of the F-theorem \cite{Klebanov:2011gs} (see also \cite{Fluder:2020pym} for further support for the conjectural 5d F-theorem).\par
As we will show in the next section, \eqref{FEFTsolved} matches with the holographic central charge \eqref{eq:doublerankchol5}, obtained from a supergravity computation.

\section{Type IIB backgrounds for 5d long quiver SCFTs}
\label{sec:gravity}
In this section we discuss the holographic dual to five-dimensional ${\cal N}=1$ SCFTs and their relevant deformations. A Poisson equation describing this dynamics is written and put in correspondence with \eqref{spe}. We give the explicit dictionary between the field theory discussion of the previous section and the holographic description below. We then propose the correspondence between ${\cal F}_{\text{EFT}}$ in \eqref{FEFTdef} and the holographic central charge, and also discuss the supergravity version of the F-theorem in \eqref{Fthm}.

\subsection{Supergravity as an electrostatic problem}\label{subsec2}
We present an infinite family of supergravity backgrounds in Type IIB string theory of the form AdS$_6 \times \mathbb{S}^2 \times \Sigma $. The isometries of AdS$_6$ correspond to the five-dimensional conformal algebra, whereas the additional $SU(2)$ isometry guaranteed by the presence of the sphere becomes the R-symmetry of the dual field theory. The 5d $\mathcal{N}=1$ SCFTs introduced in \eqref{fig:quiverfiga} live on the asymptotic boundary of AdS$_6$, with the specifics of the quiver encoded in the dependence of the metric and additional fields on the coordinates $(\sigma, \eta)$, that define the auxiliary two-dimensional geometry $\Sigma$.\par
The full string theory configuration consists of a metric, dilaton $\Phi$, $B_2$ field in the Neveu--Schwarz sector and $C_2$ and $C_0$ fields in the Ramond sector. The type IIB background in string frame is \cite{Legramandi:2021uds} (see also \cite{Apruzzi:2018cvq})
\begin{equation}\label{backgroundrescaled}
\begin{aligned}
	ds_{10}^2& = f_1 \left[ ds^2_{\text{AdS}_6} + f_2 ds^2_{\mathbb{S}^2} + f_3 (d\sigma^2+d\eta^2) \right]  , \\
e^{-2\Phi} &=f_6, \quad B_2=f_4 \text{Vol}(\mathbb{S}^2), \\ 
C_2 &= f_5 \text{Vol}(\mathbb{S}^2),\qquad  C_0= f_7 .
\end{aligned}
\end{equation}
The warp factors $f_{i} = f_i (\sigma, \eta)$ have been obtained in \cite{Legramandi:2021uds} and we report them in the appendix. They depend only on the coordinates $(\sigma, \eta)$ on $\Sigma$, and carry functional dependence on a single potential function $V_5(\sigma,\eta)$. The configuration is fixed by demanding that $V_5$ solves the linear partial differential equation (PDE)
\begin{equation}
\partial_\sigma \left(\sigma^2 \partial_\sigma V_5\right) +\sigma^2 \partial^2_\eta V_5=0.\label{diffeq5}
\end{equation}
This equation is equivalent to the BPS equations of the system.\par
The infinite family of backgrounds \eqref{backgroundrescaled}-\eqref{diffeq5} has been proven to be in exact correspondence with the solutions discussed in \cite{DHoker:2016ujz,DHoker:2016ysh,DHoker:2017mds,DHoker:2017zwj}. Moreover, the paper \cite{Legramandi:2021uds} presents the study of the  PDE \eqref{diffeq5}, with suitable boundary conditions leading to a proper interpretation of the solutions, with quantised Page charges and avoiding badly-singular behaviours.
Following \cite{Legramandi:2021uds}, we make the redefinition
\begin{equation}\label{eq:V5toWhat}
	V_5(\sigma,\eta)=\frac{\hat{W} (\sigma,\eta)}{\sigma} . 
\end{equation}
The PDE \eqref{diffeq5} then reads like a Laplace equation in flat space,
\begin{equation} \label{eq:5dLaplace}
\partial^2_\sigma \hat{W}+ \partial_\eta^2 \hat{W}=0.
\end{equation}
We stress that \eqref{eq:5dLaplace} implies that the Einstein, Maxwell and dilaton equations are satisfied. Away from the sources, the Bianchi identities are also satisfied. In the presence of sources (that play the role of flavour branes, realising the flavour groups of the dual field theory), the differential equation gets an inhomogeneous term. In that case, \eqref{eq:5dLaplace} is modified to 
\begin{equation} \label{eq:rankpoisson}
    \partial_\eta^2\hat{W} + \partial_\sigma^2\hat{W}+  \mathcal{R} ( \eta) \delta(\sigma)\,=0 ,
\end{equation}
on which we impose boundary conditions
\begin{equation}\label{bcW}
    \hat{W} (\sigma \rightarrow \pm \infty) = \hat{W}(\eta=0) =\hat{W}(\eta = P) = 0\,.
\end{equation}
This is a Poisson equation for  a classical electrostatic potential in the presence of the charge distribution encoded by $ \mathcal{R} ( \eta)$. The latter can be put in correspondence with a linear quiver describing the low energy dynamics of the five dimensional SCFT deformed by a relevant operator. Intuitively, the $\eta$-direction is a \textit{field theory} direction, moving along the quiver. The charge distribution $\mathcal{R} (\eta)$ is then identified with the rank function of Section \ref{sec:qft}. See \cite{Legramandi:2021uds} for the details.\par
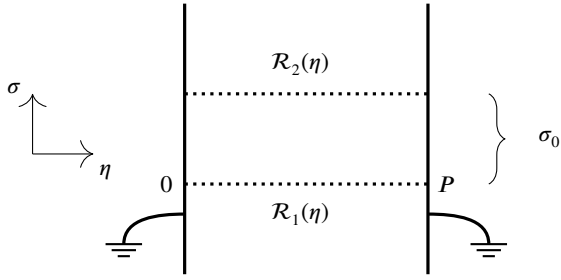
\begin{figure}
	\centering
	\begin{tikzpicture}[scale=0.8]
	\draw[very thick] (0,0) -- (0,4.5); 
	\draw[very thick] (4,0) -- (4,4.5);
	\draw[very thick] (0,1) to [out=-180,in=90] (-1,0.5);
	\draw[very thick] (4,1) to [out=0,in=90] (5,0.5);
	\draw[thick] (4.7,0.5) -- (5.3,0.5);
	\draw[thick] (4.8,0.4) -- (5.2,0.4);
	\draw[thick] (4.9,0.3) -- (5.1,0.3);
	\draw[thick] (-.7,0.5) -- (-1.3,0.5);
	\draw[thick] (-.8,0.4) -- (-1.2,0.4);
	\draw[thick] (-.9,0.3) -- (-1.1,0.3);
	\draw[->] (-2.5,2) -- (-2.5,3);
	\draw[->] (-2.5,2) -- (-1.5,2);
	\node at (-2.8,3.1) {\small $\sigma$};
	\node at (-1.3,1.7) {\small $\eta$};
	\node at (6,2.25) {$\sigma_0$};
	\draw[very thick,dotted] (0,1.5) -- (4,1.5); 
	\node at (-0.3,1.5) {$0$};
	\node at (4.3,1.5) {$P$};
	\node at (1.9,1) {$\mathcal{R}_1(\eta)$};
	
	\draw[very thick,dotted] (0,3) -- (4,3); 
	\node at (1.9,3.5) {$\mathcal{R}_2(\eta)$};
	
	\draw [decorate,
    decoration = {brace, mirror, amplitude=6pt}] (5,1.5) --  (5,3);

	\end{tikzpicture}
	\caption{The electrostatic problem for $\hat{W}$ in the two rank functions setup. A charge distribution equal to $\mathcal{ R}_1 (\eta)$ is placed at $\sigma=0$ and a second charge distribution $\mathcal{R}_2(\eta)$ is at $\sigma=\sigma_0$.}
	\label{fig:doublerankelectrostatic}
\end{figure}\par
In this letter we are interested in exploring the physical significance of the $\sigma$-direction. To do this, we generalise the electrostatic problem and consider the situation in which two charge densities $ \mathcal{R}_1 ( \eta) $ and $ \mathcal{R}_2 ( \eta) $ are separated in  $\sigma$ by a distance $\sigma_0$. In this setup, the Poisson equation \eqref{eq:rankpoisson} encoding the dynamics of the system reads 
\begin{equation} \label{eq:doublerankpoisson}
    \partial_\eta^2\hat{W}(\sigma, \eta) + \partial_\sigma^2\hat{W}(\sigma, \eta) + \mathcal{R}_1(\eta)\delta(\sigma) + \mathcal{R}_2 ( \eta) \delta(\sigma-\sigma_0)\,=0 ,
\end{equation}
with boundary conditions \eqref{bcW}. The associated electrostatic problem is depicted in Fig.~\ref{fig:doublerankelectrostatic}. The  solution is \cite{Akhond:2022oaf}
\begin{equation} \label{eq:doublerankWhat}
    \hat{W} = \sum_{k=1}^\infty \frac{P}{2 k \pi} \sin\left(\frac{k \pi \eta}{P}\right)\left[R_{1,k} e^{-\frac{k \pi}{P}|\sigma|} + R_{2,k} e^{-\frac{k \pi}{P}|\sigma-\sigma_0|}\right]\,,
\end{equation}
with $R_{1,k}$ and $R_{2,k}$ the Fourier coefficients of the two rank functions, as in \eqref{eq:doublefourier}.\par
The electrostatic problem is directly translated in the saddle point problem of Section \ref{sec:qft}. Indeed, dividing \eqref{eq:doublerankpoisson} by $-3N/2$ and acting with $\partial_{\eta}^2$, 
we reproduce \eqref{spe} upon identification  
\begin{equation}
\label{holotocft}
	\phi = \frac{3}{2} \sigma , \quad m = \frac{3}{2} \sigma_0 , \quad -\partial_{\eta}^2 \hat{W}  = \frac{3N}{2}\rho  ,
\end{equation}
while $\eta$ is mapped trivially.

\subsection{Holographic central charge}
An interesting observable that informs us about the interpretation of the parameter $\sigma_0$ is the holographic central charge $c_{hol}$. See \cite{Macpherson:2014eza, Bea:2015fja} for a general definition and \cite{Legramandi:2021uds} for the calculation in the context of AdS$_6$ backgrounds. $c_{hol}$ is proportional to the free energy calculated in the previous section. More precisely, at the SCFT point $\sigma_0 \to 0$, $\frac{27\pi^6}{4} c_{hol}$ equals the limit $m \to 0$ of \eqref{F}. We now proceed to compute the holographic central charge at arbitrary $\sigma_0$ and show that 
\begin{equation}
	\frac{27\pi^6}{4} c_{hol} = \mf_{\text{EFT}} .
\end{equation}\par
Intuitively, the holographic central charge of the theory is computed as the weighted-volume of the internal manifold $\mathbb{S}^2 \times \Sigma $. After a lengthy calculation, this works out to be \cite{Legramandi:2021uds}
\begin{equation}
    c_{hol} = \frac{2}{3 \pi^5} \int_{-\infty}^\infty d\sigma\int_0^Pd\eta\, \partial^2_\eta \hat{W}\left( \sigma\partial_\sigma \hat{W}- \hat{W}\right).  
\label{eq:chol5dV}
\end{equation}
Here $\hat{W}$ is given by \eqref{eq:doublerankWhat} and the residual integration is over the surface $\Sigma$. After the integrals are performed, one finds 
\begin{equation} \label{eq:doublerankchol5}
\begin{aligned}
     c_{hol} [\mathcal{R}_1, \mathcal{R}_2] & = \frac{P^2}{8 \pi^6} \sum_{k=1}^\infty \frac{1}{k}\left[ R_{1,k}^2 + R_{2,k}^2 \right. \\
     & \left. + 2 R_{1,k}R_{2,k} e^{-\frac{k \pi \sigma_0}{P}} \left( 1+ \frac{k \pi \sigma_0}{P} \right) \right]\,.
\end{aligned}
\end{equation}
With the identification $\frac{\sigma_0}{2} = \frac{m}{3}$ already found in \eqref{holotocft}, this is proportional to \eqref{FEFTsolved}, as claimed.\par
In the spirit of a holographic version of the F-theorem, we state here that $c_{hol}$ satisfies 
\begin{equation}
\begin{aligned}
	\lim_{\sigma_0/P \to 0} c_{hol}[\mathcal{R}_1, \mathcal{R}_2] &= c_{hol}[\mathcal{R}_1 +\mathcal{R}_2] , \\
	 \lim_{\sigma_0/P \to \infty} c_{hol}[\mathcal{R}_1, \mathcal{R}_2] &= c_{hol} \left[ \mathcal{R}_1 \right] + c_{hol} \left[ \mathcal{R}_2 \right] \\
		\lim_{\sigma_0/P \to 0} c_{hol} & > \lim_{\sigma_0/P \to \infty} c_{hol} \\
	\partial_{\sigma_0}c_{hol} &\le 0 
\end{aligned}\label{diego}
\end{equation}
This mirrors the statement \eqref{Fthm}. It is shown rigorously in \cite{Akhond:2022oaf} that the F-theorem is valid for this kind of relevant deformations. The proof goes in two parts. On the field theory side, it shows that $\mf_{\text{EFT}} $ in \eqref{FEFTsolved} satisfies the conditions in \eqref{Fthm}. On the other hand, it is shown independently that the holographic central charge \eqref{eq:doublerankchol5} satisfies \eqref{diego}.\par
\medskip
While $c_{hol}$ (as well as probe branes \cite{Akhond:2022oaf}) clearly see the factorisation, only one metric \eqref{backgroundrescaled} remains as $\sigma_0 \to \infty$. However, depending on how the limit is taken, either the metric given by $\mathcal{R}_1$ or the one given by $\mathcal{R}_2$ is retained. Namely, sending $\sigma_0 \to \infty$ directly, one is left with the AdS$_6$ metric fixed by the rank function $\mathcal{R}_1$, as in \cite{Legramandi:2021uds}. However, first shifting $\tilde{\sigma} = \sigma - \sigma_0$ and then sending $\sigma_0 \to \infty$, the AdS$_6$ metric that remains is determined by the rank function $\mathcal{R}_2$, located at $\tilde{\sigma}=0$. Therefore, even if only one metric survives at the end of the RG flow, the geometry still detects the existence of two decoupled SCFTs. The choice of how to take the limit, and thus of which metric retain in the IR, is precisely dual to the choice of conformal vacuum in the RG flow of the dual field theory.

\subsection{Study of singularities}
Let us turn our attention to the analysis of the singularities in the backgrounds \eqref{backgroundrescaled}, once $V_5$ in \eqref{eq:V5toWhat}-\eqref{eq:doublerankWhat} is used. Already for $\sigma_0=0$, the backgrounds \eqref{backgroundrescaled} do present singularities that, in the regime of very large $N,P$, are sharply localised at the location of the source branes in the Type IIB setup. The result is reliable everywhere away from the sources, although very close to them one should supplement the analysis with the open string sector.\par
Nevertheless, for finite $\sigma_0$, new singularities appear, which cannot be interpreted in terms of localised sources. The supergravity backgrounds are not a faithful dual description of the field theory dynamics, because the parameter $\sigma_0$ introduces a finite scale and is incompatible with conformal invariance of the field theory, and thus with the AdS$_6$ factor in the family of metrics. The supergravity solution we would need to construct deforms away from AdS$_6$ into a $\R^{1,4}$ and a radial coordinate, which should be a combination of the AdS radial coordinate and the $\sigma$-direction. For
$|\sigma_0|\to\infty$ the AdS$_6$ factor is eventually recovered, with a factorised form of the warp factors.\par
The behaviour of these more elaborated geometries should then be in correspondence with the SCFT deformed by a relevant operator, with parameter proportional to $\sigma_0$, that flows into two decoupled SCFTs as indicated in Fig.~\ref{fig:flow}. A background fluctuation localised close to $\sigma =0$ should have fast decay for $|\sigma|\to\infty$. We leave these explicit checks for future investigation.\par
It is natural to wonder about the validity of $c_{hol}$ in \eqref{eq:doublerankchol5}, as it is obtained using the background \eqref{backgroundrescaled} with potential \eqref{eq:doublerankWhat}. Carefully going over the supergravity calculation shows that dilaton and
the metric components that present a singular behaviour, do in fact cancel, yielding the final result \eqref{eq:chol5dV}-\eqref{eq:doublerankchol5}. The same cancellations occur in other observables \cite{Akhond:2022oaf}, and are in line with closely related early observations in \cite{Gutperle:2017tjo}. Even when the singular behaviour of our backgrounds are different, this type of cancellations are reminiscent of the so-called \textit{good singularities} \cite{Gubser:2000nd,Maldacena:2000mw}. It must be emphasised that other observables will display a singular behaviour for finite $\sigma_0$, invalidating the AdS$_6$ ansatz \eqref{backgroundrescaled} as a description of the mass-deformed SCFT.

\section{Conclusions}

We have studied RG flows associated to relevant deformations of long quivers. Insisting on the class of mass deformations that yield balanced linear quiver SCFTs both at the beginning and end of the RG flow, we have explored the mass deformations that break the SCFT in two (or more) linear quivers. After understanding these RG flows from the perspective of the Coulomb branch, we have taken the large $N$ limit and computed the free energy at arbitrary points of the RG flow. We have also discussed the holographic dual of such deformations, and given a neat interpretation in terms on an electrostatic problem.\par
The main findings reported in this letter are:
\begin{itemize}
	\item On the QFT side, we introduced the quantity $\mf_{\text{EFT}}$ in \eqref{FEFTdef}, tailored to probe the RG flows between interacting SCFTs. It depends on the mass parameter, but not on the choice of vacuum. We have evaluated it in \eqref{FEFTsolved} and shown that it satisfies the F-theorem.
	\item On the AdS$_6$ side, we have mapped the supergravity problem to an electrostatic one. We have then identified the holographic dual of the mass deformation in this electrostatic formalism. We have used the resulting solutions to compute $c_{hol}$, finding agreement with $\mf_{\text{EFT}}$.
\end{itemize}
We have thus provided a window on the enhancement of the AdS/CFT framework to entire RG flows connecting 5d $\mathcal{N}=1$ SCFTs. Having identified the appropriate quantity for such scope, we have also used it to support the 5d F-theorem.\par
The electrostatic formalism points toward several generalisations. Understanding their physical significance in supergravity on the one hand, and identifying and studying the dual deformations of 5d $\mathcal{N}=1$ SCFTs on the other, constitute a long term project to enrich the holographic dictionary.

\section*{Acknowledgments}
The contents and presentation of this work much benefited from extensive discussion with various colleagues. We would like to specially thank: Lorenzo Coccia, Diego Correa, Stefano Cremonesi, Michele Del Zotto, Diego Rodriguez-Gomez, Mauricio Romo, Masazumi Honda, Yolanda Lozano, Ali Mollabashi, Matteo Sacchi, Guillermo Silva, Daniel Thompson, Miguel Tierz, Alessandro Tomasiello, Christoph Uhlemann, Yifan Wang.
AL and CN are supported by STFC grant ST/T000813/1. LS is supported by The Royal Society through grant RGF\textbackslash R1\textbackslash 180087 Generalised Dualities, Resurgence and Integrability.  
AL has also received funding from the European Research Council (ERC) under the European Union's Horizon 2020 research and innovation programme (grant agreement No 804305).
For the purpose of open access, the authors have applied a Creative Commons Attribution (CC BY) licence to any Author Accepted Manuscript version arising.

\section*{Appendix}
The warp factors $f_1, \dots , f_7$ appearing in the supergravity background \eqref{backgroundrescaled} are:
\begin{align*}\label{backgroundrescaled-c}
f_1&= \frac{3 \pi}{2}\sqrt{\sigma^2 +\frac{3\sigma \partial_\sigma V_5}{\partial^2_\eta V_5}} ,\\
f_2&= \frac{\partial_\sigma V_5 \partial^2_\eta V_5}{3\Lambda},\\
f_3&= \frac{\partial^2_\eta V_5}{3\sigma \partial_\sigma V_5}, \\
f_4&= \frac{\pi}{2}\left[\eta -\frac{ \sigma \partial_\sigma V_5 \partial_\sigma\partial_\eta V_5 }{\Lambda} \right] ,\\
f_5&= \frac{\pi}{2}\left[ V_5 - \frac{\sigma\partial_\sigma V_5}{\Lambda} \left( \partial_\eta V_5  \partial_\sigma \partial_\eta V_5  -3 \partial^2_\eta V_5  \partial_\sigma V_5 \right) \right] ,\\
f_6&= \frac{12  \Lambda \left( \sigma^2 \partial_\sigma V_5 \partial^2_\eta V_5 \right)}{\left(3 \partial_\sigma V_5 +\sigma \partial^2_\eta V_5 \right)^2} ,\\ 
f_7&= 2\left[ \partial_\eta V_5 + \frac{ 3\sigma \partial_\sigma V_5  \partial_\sigma\partial_\eta V_5 }{3\partial_\sigma V_5 +\sigma \partial^2_\eta V_5}  \right] , 
\end{align*}
and the function $\Lambda$ is defined as 
\begin{equation*}
	\Lambda = \sigma \left( \partial_\sigma\partial_\eta V_5 \right)^2 + \left( \partial_\sigma V_5-\sigma \partial^2_\sigma V_5 \right)  \partial^2_\eta V_5 .
\end{equation*}
The map from \cite{DHoker:2016ujz,DHoker:2016ysh,DHoker:2017mds,DHoker:2017zwj} to the present setup requires fixing the origin in the $(\sigma, \eta)$-plane \cite{Legramandi:2021uds}. These expressions are written assuming a rank function located at $\sigma=0$, and possibly adding other rank functions. In the case of a single rank function located at $\sigma= \sigma_0$, one should first shift $\tilde{\sigma}= \sigma-\sigma_0$ and then replace $\sigma$ with $\tilde{\sigma}$ in the expressions.

\bibliographystyle{mystyle}

\bibliography{holo5d3d}

\providecommand{\href}[2]{#2}\begingroup\raggedright\begin{thebibliography}{10}

\bibitem{Akhond:2022oaf}
M.~Akhond, A.~Legramandi, C.~Nunez, L.~Santilli and L.~Schepers, ``{Matrix
  Models and Holography: Mass Deformations of Long Quiver Theories in 5d and
  3d}.''\href{https://doi.org/10.21468/SciPostPhys.15.3.086}{\emph{SciPost
  Phys.} {\bfseries 15} (2023) 086}
  [\href{https://arxiv.org/abs/2211.13240}{{\ttfamily 2211.13240}}].

\bibitem{Nahm:1977tg}
W.~Nahm, ``{Supersymmetries and their
  Representations}.''\href{https://doi.org/10.1016/0550-3213(78)90218-3}{\emph{Nucl.
  Phys. B} {\bfseries 135} (1978) 149}.

\bibitem{Seiberg:1996bd}
N.~Seiberg, ``{Five-dimensional SUSY field theories, nontrivial fixed points
  and string
  dynamics}.''\href{https://doi.org/10.1016/S0370-2693(96)01215-4}{\emph{Phys.
  Lett. B} {\bfseries 388} (1996) 753}
  [\href{https://arxiv.org/abs/hep-th/9608111}{{\ttfamily hep-th/9608111}}].

\bibitem{Cordova:2016xhm}
C.~Cordova, T.~T. Dumitrescu and K.~Intriligator, ``{Deformations of
  Superconformal
  Theories}.''\href{https://doi.org/10.1007/JHEP11(2016)135}{\emph{JHEP}
  {\bfseries 11} (2016) 135}
  [\href{https://arxiv.org/abs/1602.01217}{{\ttfamily 1602.01217}}].

\bibitem{Chang:2018xmx}
C.-M. Chang, ``{5d and 6d SCFTs Have No Weak Coupling
  Limit}.''\href{https://doi.org/10.1007/JHEP09(2019)016}{\emph{JHEP}
  {\bfseries 09} (2019) 016}
  [\href{https://arxiv.org/abs/1810.04169}{{\ttfamily 1810.04169}}].

\bibitem{Aharony:1997bh}
O.~Aharony, A.~Hanany and B.~Kol, ``{Webs of (p,q) five-branes,
  five-dimensional field theories and grid
  diagrams}.''\href{https://doi.org/10.1088/1126-6708/1998/01/002}{\emph{JHEP}
  {\bfseries 01} (1998) 002}
  [\href{https://arxiv.org/abs/hep-th/9710116}{{\ttfamily hep-th/9710116}}].

\bibitem{Intriligator:1997pq}
K.~A. Intriligator, D.~R. Morrison and N.~Seiberg, ``{Five-dimensional
  supersymmetric gauge theories and degenerations of Calabi-Yau
  spaces}.''\href{https://doi.org/10.1016/S0550-3213(97)00279-4}{\emph{Nucl.
  Phys. B} {\bfseries 497} (1997) 56}
  [\href{https://arxiv.org/abs/hep-th/9702198}{{\ttfamily hep-th/9702198}}].

\bibitem{Pestun:2007rz}
V.~Pestun, ``{Localization of gauge theory on a four-sphere and supersymmetric
  Wilson
  loops}.''\href{https://doi.org/10.1007/s00220-012-1485-0}{\emph{Commun. Math.
  Phys.} {\bfseries 313} (2012) 71}
  [\href{https://arxiv.org/abs/0712.2824}{{\ttfamily 0712.2824}}].

\bibitem{Maldacena:1997re}
J.~M. Maldacena, ``{The Large N limit of superconformal field theories and
  supergravity}.''\href{https://doi.org/10.1023/A:1026654312961}{\emph{Adv.
  Theor. Math. Phys.} {\bfseries 2} (1998) 231}
  [\href{https://arxiv.org/abs/hep-th/9711200}{{\ttfamily hep-th/9711200}}].

\bibitem{Jafferis:2012iv}
D.~L. Jafferis and S.~S. Pufu, ``{Exact results for five-dimensional
  superconformal field theories with gravity
  duals}.''\href{https://doi.org/10.1007/JHEP05(2014)032}{\emph{JHEP}
  {\bfseries 05} (2014) 032} [\href{https://arxiv.org/abs/1207.4359}{{\ttfamily
  1207.4359}}].

\bibitem{Alday:2014rxa}
L.~F. Alday, M.~Fluder, P.~Richmond and J.~Sparks, ``{Gravity Dual of
  Supersymmetric Gauge Theories on a Squashed
  Five-Sphere}.''\href{https://doi.org/10.1103/PhysRevLett.113.141601}{\emph{Phys.
  Rev. Lett.} {\bfseries 113} (2014) 141601}
  [\href{https://arxiv.org/abs/1404.1925}{{\ttfamily 1404.1925}}].

\bibitem{Alday:2014bta}
L.~F. Alday, M.~Fluder, C.~M. Gregory, P.~Richmond and J.~Sparks,
  ``{Supersymmetric gauge theories on squashed five-spheres and their gravity
  duals}.''\href{https://doi.org/10.1007/JHEP09(2014)067}{\emph{JHEP}
  {\bfseries 09} (2014) 067} [\href{https://arxiv.org/abs/1405.7194}{{\ttfamily
  1405.7194}}].

\bibitem{Chang:2017mxc}
C.-M. Chang, M.~Fluder, Y.-H. Lin and Y.~Wang, ``{Romans Supergravity from
  Five-Dimensional
  Holograms}.''\href{https://doi.org/10.1007/JHEP05(2018)039}{\emph{JHEP}
  {\bfseries 05} (2018) 039}
  [\href{https://arxiv.org/abs/1712.10313}{{\ttfamily 1712.10313}}].

\bibitem{Romans:1985tw}
L.~J. Romans, ``{The F(4) Gauged Supergravity in
  Six-dimensions}.''\href{https://doi.org/10.1016/0550-3213(86)90517-1}{\emph{Nucl.
  Phys. B} {\bfseries 269} (1986) 691}.

\bibitem{Cvetic:1999un}
M.~Cvetic, H.~Lu and C.~N. Pope, ``{Gauged six-dimensional supergravity from
  massive type
  IIA}.''\href{https://doi.org/10.1103/PhysRevLett.83.5226}{\emph{Phys. Rev.
  Lett.} {\bfseries 83} (1999) 5226}
  [\href{https://arxiv.org/abs/hep-th/9906221}{{\ttfamily hep-th/9906221}}].

\bibitem{Bergman:2012kr}
O.~Bergman and D.~Rodriguez-Gomez, ``{5d quivers and their AdS(6)
  duals}.''\href{https://doi.org/10.1007/JHEP07(2012)171}{\emph{JHEP}
  {\bfseries 07} (2012) 171} [\href{https://arxiv.org/abs/1206.3503}{{\ttfamily
  1206.3503}}].

\bibitem{DHoker:2016ujz}
E.~D'Hoker, M.~Gutperle, A.~Karch and C.~F. Uhlemann, ``{Warped $AdS_6\times
  S^2$ in Type IIB supergravity I: Local
  solutions}.''\href{https://doi.org/10.1007/JHEP08(2016)046}{\emph{JHEP}
  {\bfseries 08} (2016) 046}
  [\href{https://arxiv.org/abs/1606.01254}{{\ttfamily 1606.01254}}].

\bibitem{DHoker:2016ysh}
E.~D'Hoker, M.~Gutperle and C.~F. Uhlemann, ``{Holographic duals for
  five-dimensional superconformal quantum field
  theories}.''\href{https://doi.org/10.1103/PhysRevLett.118.101601}{\emph{Phys.\
  Rev.\ Lett.} {\bfseries 118} (2017) 101601}
  [\href{https://arxiv.org/abs/1611.09411}{{\ttfamily 1611.09411}}].

\bibitem{DHoker:2017mds}
E.~D'Hoker, M.~Gutperle and C.~F. Uhlemann, ``{Warped $AdS_6\times S^2$ in Type
  IIB supergravity II: Global solutions and five-brane
  webs}.''\href{https://doi.org/10.1007/JHEP05(2017)131}{\emph{JHEP} {\bfseries
  05} (2017) 131} [\href{https://arxiv.org/abs/1703.08186}{{\ttfamily
  1703.08186}}].

\bibitem{DHoker:2017zwj}
E.~D'Hoker, M.~Gutperle and C.~F. Uhlemann, ``{Warped $AdS_6\times S^2$ in Type
  IIB supergravity III: Global solutions with
  seven-branes}.''\href{https://doi.org/10.1007/JHEP11(2017)200}{\emph{JHEP}
  {\bfseries 11} (2017) 200}
  [\href{https://arxiv.org/abs/1706.00433}{{\ttfamily 1706.00433}}].

\bibitem{Gutperle:2018axv}
M.~Gutperle, J.~Kaidi and H.~Raj, ``{Mass deformations of 5d SCFTs via
  holography}.''\href{https://doi.org/10.1007/JHEP02(2018)165}{\emph{JHEP}
  {\bfseries 02} (2018) 165}
  [\href{https://arxiv.org/abs/1801.00730}{{\ttfamily 1801.00730}}].

\bibitem{Gutperle:2018vdd}
M.~Gutperle, A.~Trivella and C.~F. Uhlemann, ``{Type IIB 7-branes in warped
  AdS$_{6}$: partition functions, brane webs and probe
  limit}.''\href{https://doi.org/10.1007/JHEP04(2018)135}{\emph{JHEP}
  {\bfseries 04} (2018) 135}
  [\href{https://arxiv.org/abs/1802.07274}{{\ttfamily 1802.07274}}].

\bibitem{Fluder:2018chf}
M.~Fluder and C.~F. Uhlemann, ``{Precision Test of AdS$_6$/CFT$_5$ in Type IIB
  String
  Theory}.''\href{https://doi.org/10.1103/PhysRevLett.121.171603}{\emph{Phys.
  Rev. Lett.} {\bfseries 121} (2018) 171603}
  [\href{https://arxiv.org/abs/1806.08374}{{\ttfamily 1806.08374}}].

\bibitem{Uhlemann:2019ypp}
C.~F. Uhlemann, ``{Exact results for 5d SCFTs of long quiver
  type}.''\href{https://doi.org/10.1007/JHEP11(2019)072}{\emph{JHEP} {\bfseries
  11} (2019) 072} [\href{https://arxiv.org/abs/1909.01369}{{\ttfamily
  1909.01369}}].

\bibitem{Uhlemann:2020bek}
C.~F. Uhlemann, ``{Wilson loops in 5d long quiver gauge
  theories}.''\href{https://doi.org/10.1007/JHEP09(2020)145}{\emph{JHEP}
  {\bfseries 09} (2020) 145}
  [\href{https://arxiv.org/abs/2006.01142}{{\ttfamily 2006.01142}}].

\bibitem{Chang:2017cdx}
C.-M. Chang, M.~Fluder, Y.-H. Lin and Y.~Wang, ``{Spheres, Charges, Instantons,
  and Bootstrap: A Five-Dimensional
  Odyssey}.''\href{https://doi.org/10.1007/JHEP03(2018)123}{\emph{JHEP}
  {\bfseries 03} (2018) 123}
  [\href{https://arxiv.org/abs/1710.08418}{{\ttfamily 1710.08418}}].

\bibitem{Legramandi:2021uds}
A.~Legramandi and C.~Nunez, ``{Electrostatic description of five-dimensional
  SCFTs}.''\href{https://doi.org/10.1016/j.nuclphysb.2021.115630}{\emph{Nucl.
  Phys. B} {\bfseries 974} (2022) 115630}
  [\href{https://arxiv.org/abs/2104.11240}{{\ttfamily 2104.11240}}].

\bibitem{Kallen:2012va}
J.~K\"all\'en, J.~Qiu and M.~Zabzine, ``{The perturbative partition function of
  supersymmetric 5D Yang-Mills theory with matter on the
  five-sphere}.''\href{https://doi.org/10.1007/JHEP08(2012)157}{\emph{JHEP}
  {\bfseries 08} (2012) 157} [\href{https://arxiv.org/abs/1206.6008}{{\ttfamily
  1206.6008}}].

\bibitem{Lockhart:2012vp}
G.~Lockhart and C.~Vafa, ``{Superconformal Partition Functions and
  Non-perturbative Topological
  Strings}.''\href{https://doi.org/10.1007/JHEP10(2018)051}{\emph{JHEP}
  {\bfseries 10} (2018) 051} [\href{https://arxiv.org/abs/1210.5909}{{\ttfamily
  1210.5909}}].

\bibitem{Chang:2019uag}
C.-M. Chang, M.~Fluder, Y.-H. Lin and Y.~Wang, ``{Proving the 6d Cardy Formula
  and Matching Global Gravitational
  Anomalies}.''\href{https://doi.org/10.21468/SciPostPhys.11.2.036}{\emph{SciPost
  Phys.} {\bfseries 11} (2021) 036}
  [\href{https://arxiv.org/abs/1910.10151}{{\ttfamily 1910.10151}}].

\bibitem{Klebanov:2011gs}
I.~R. Klebanov, S.~S. Pufu and B.~R. Safdi, ``{F-Theorem without
  Supersymmetry}.''\href{https://doi.org/10.1007/JHEP10(2011)038}{\emph{JHEP}
  {\bfseries 10} (2011) 038} [\href{https://arxiv.org/abs/1105.4598}{{\ttfamily
  1105.4598}}].

\bibitem{Fluder:2020pym}
M.~Fluder and C.~F. Uhlemann, ``{Evidence for a 5d
  F-theorem}.''\href{https://doi.org/10.1007/JHEP02(2021)192}{\emph{JHEP}
  {\bfseries 02} (2021) 192}
  [\href{https://arxiv.org/abs/2011.00006}{{\ttfamily 2011.00006}}].

\bibitem{Apruzzi:2018cvq}
F.~Apruzzi, J.~C. Geipel, A.~Legramandi, N.~T. Macpherson and M.~Zagermann,
  ``{Minkowski$_4$ $\times$ $S^2$ solutions of IIB
  supergravity}.''\href{https://doi.org/10.1002/prop.201800006}{\emph{Fortsch.
  Phys.} {\bfseries 66} (2018) 1800006}
  [\href{https://arxiv.org/abs/1801.00800}{{\ttfamily 1801.00800}}].

\bibitem{Macpherson:2014eza}
N.~T. Macpherson, C.~N\'u\~nez, L.~A. Pando~Zayas, V.~G.~J. Rodgers and C.~A.
  Whiting, ``{Type IIB supergravity solutions with AdS$_{5}$ from Abelian and
  non-Abelian T
  dualities}.''\href{https://doi.org/10.1007/JHEP02(2015)040}{\emph{JHEP}
  {\bfseries 02} (2015) 040} [\href{https://arxiv.org/abs/1410.2650}{{\ttfamily
  1410.2650}}].

\bibitem{Bea:2015fja}
Y.~Bea, J.~D. Edelstein, G.~Itsios, K.~S. Kooner, C.~Nunez, D.~Schofield
  et~al., ``{Compactifications of the Klebanov-Witten CFT and new AdS$_{3}$
  backgrounds}.''\href{https://doi.org/10.1007/JHEP05(2015)062}{\emph{JHEP}
  {\bfseries 05} (2015) 062}
  [\href{https://arxiv.org/abs/1503.07527}{{\ttfamily 1503.07527}}].

\bibitem{Gutperle:2017tjo}
M.~Gutperle, C.~Marasinou, A.~Trivella and C.~F. Uhlemann, ``{Entanglement
  entropy vs. free energy in IIB supergravity duals for 5d
  SCFTs}.''\href{https://doi.org/10.1007/JHEP09(2017)125}{\emph{JHEP}
  {\bfseries 09} (2017) 125}
  [\href{https://arxiv.org/abs/1705.01561}{{\ttfamily 1705.01561}}].

\bibitem{Gubser:2000nd}
S.~S. Gubser, ``{Curvature singularities: The Good, the bad, and the
  naked}.''\href{https://doi.org/10.4310/ATMP.2000.v4.n3.a6}{\emph{Adv. Theor.
  Math. Phys.} {\bfseries 4} (2000) 679}
  [\href{https://arxiv.org/abs/hep-th/0002160}{{\ttfamily hep-th/0002160}}].

\bibitem{Maldacena:2000mw}
J.~M. Maldacena and C.~Nunez, ``{Supergravity description of field theories on
  curved manifolds and a no go
  theorem}.''\href{https://doi.org/10.1142/S0217751X01003937}{\emph{Int. J.
  Mod. Phys. A} {\bfseries 16} (2001) 822}
  [\href{https://arxiv.org/abs/hep-th/0007018}{{\ttfamily hep-th/0007018}}].

\end{thebibliography}\endgroup

\end{document}